\documentclass{article}

\usepackage[preprint, nonatbib]{neurips_2024}
\usepackage[numbers]{natbib}

\usepackage[utf8]{inputenc} 
\usepackage[T1]{fontenc}    
\usepackage{hyperref}       
\usepackage{url}            
\usepackage{booktabs}       
\usepackage{amsfonts}       
\usepackage{nicefrac}       
\usepackage{microtype}      
\usepackage{xcolor}         

\usepackage{amsmath}
\usepackage{graphicx}
\usepackage{adjustbox}

\usepackage{caption}
\usepackage{cleveref}

\DeclareCaptionLabelFormat{suppfigure}{Supplemental Figure S#2}
\DeclareCaptionLabelFormat{supptable}{Supplemental Table S#2}

\title{Integrating GNN and Neural ODEs for Estimating Non-Reciprocal Two-Body Interactions in Mixed-Species Collective Motion}

\newcommand{\TODO}[1]{#1}
\newcommand{\SKS}[1]{#1}
\newcommand{\MU}[1]{#1}

\author{Masahito Uwamichi\(^{1*}\), Simon K. Schnyder\(^{2}\), Tetsuya J. Kobayashi\(^{2,3}\), Satoshi Sawai\(^{1,3}\)\\
\(^{1}\)Graduate School of Arts and Sciences, The University of Tokyo\\
\(^{2}\)Institute of Industrial Science, The University of Tokyo\\
\(^{3}\)Universal Biology Institute, The University of Tokyo\\
\texttt{uwamichi@g.ecc.u-tokyo.ac.jp},~\texttt{simon@sat.t.u-tokyo.ac.jp},\\
\texttt{tetsuya@sat.t.u-tokyo.ac.jp},~\texttt{cssawai@mail.ecc.u-tokyo.ac.jp}}

\begin{document}

\maketitle

\begin{abstract}
Analyzing the motion of multiple biological agents, be it cells or individual animals, is pivotal for the understanding of complex collective behaviors. 
With the advent of advanced microscopy, detailed images of complex tissue formations involving multiple cell types have become more accessible in recent years. 
However, deciphering the underlying rules that govern cell movements is far from trivial. 
Here, we present a novel deep learning framework for estimating the underlying equations of motion from observed trajectories, a pivotal step in decoding such complex dynamics. 
Our framework integrates graph neural networks with neural differential equations, enabling effective prediction of two-body interactions based on the states of the interacting entities. 
We demonstrate the efficacy of our approach through two numerical experiments. 
First, we used simulated data from a toy model to tune the hyperparameters. 
Based on the obtained hyperparameters, we then applied this approach to a more complex model with non-reciprocal forces that mimic the collective dynamics of the cells of slime molds. 
Our results show that the proposed method can accurately estimate the functional forms of two-body interactions -- even when they are nonreciprocal -- thereby precisely replicating both individual and collective behaviors within these systems.
\end{abstract}

\section{Introduction}

Collective motion, a phenomenon observed in various biological systems, is characterized by the coordinated movement of multiple entities. 
This behavior is prevalent in a wide range of self-propelled systems, collectively referred to as active matter~\cite{Vicsek2012}, from flocks of birds~\cite{Cavagna2010,flack2018local,Charlesworth2019} and schools of fish~\cite{Katz2011} to cellular slime molds~\cite{fujimori2019}, microswimmers~\cite{Elgeti2015}, \TODO{swarming leukocytes~\cite{strickland2024},} and human crowds~\cite{Kok2016}. 
Understanding the underlying mechanisms of collective motion is crucial for elucidating the principles governing the dynamics of these systems. 
In particular, the interactions between individual entities play a pivotal role in shaping the collective behavior of the system. 

Recent advances in imaging technologies have enabled detailed observations at the cellular level, providing insights into the dynamics of complex tissue formation involving multiple cell types. 
For example, cellular slime molds, a model organism for studying collective motion, exhibit intricate behaviors such as aggregation, migration, and differentiation~\cite{fujimori2019}. 
These behaviors are driven by the interactions between different cell types, which are mediated by chemical signals and physical forces. 
Decoding the underlying equations of motion that govern these interactions is essential for understanding the emergent properties of these systems.

In this work, we present a novel deep learning framework for estimating two-body interactions in a mixed-species collective motion. 
Our framework integrates graph neural networks (GNNs) with neural differential equations (neural DEs) to predict interactions between pairs of entities based on their states. 
GNNs are well-suited for modeling complex interactions in graph-structured data, while neural DEs provide a flexible framework for learning the dynamics of the system. 
By combining these two approaches, we can effectively capture the interactions between individual entities and predict their collective behavior.
We demonstrate the efficacy of our framework through two numerical experiments. 
The first experiment uses a toy model designed to generate data for refining the hyperparameters of our framework.
The second experiment explores a complex scenario \MU{that partially mimics} the collective motion of cellular slime molds, where two different cell types interact with each other. 
Our results show that our method can accurately estimate the two-body interactions, thereby replicating both individual and collective behaviors within these systems.

The rest of this paper is organized as follows.
In \cref{sec:background}, we introduce the study of collective motion.
In \cref{sec:relatedwork}, we provide an overview of related work on collective motion and deep learning for dynamical systems. 
In \cref{sec:method}, we describe our deep learning framework for estimating two-body interactions in mixed-species collective motion. 
In \cref{sec:Exp}, we successfully apply this framework to two numerical experiments. 
Finally, in \cref{sec:conclusion}, we discuss the implications of our work and outline potential future research directions.

\section{Background}\label{sec:background}

Let us first describe the formulation of collective motion in active matter. Starting with the Vicsek model~\cite{Vicsek1995}, the collective behavior in active matter is described based on the centroids, velocities, and orientations of each individual. A unique aspect of active matter is that it allows for spontaneous generation of forces and torques, which is justified by the ability to extract and utilize energy from the external environment~\cite{Menon2010}.
The Vicsek model itself is a multi-particle model in discrete time, where each individual possesses a velocity along its orientation and adjusts its direction based on interactions with nearby individuals. This model assumes that interactions among individuals are local and can be represented using a \MU{dynamical} graph structure~\cite{boccaletti2006}. Stemming from this model, numerous other models have been proposed, differing in the nature of interactions between individuals and the forms of their equations of motion.
Particularly models with continuous-time motion equations often use Langevin equations where the motion of individuals is described by the summation of pairwise interactions~\cite{chate2008collective}. Furthermore, since active matter can utilize energy from the external environment, focusing solely on moving agents categorizes it as an open system. This implies that the conservation of the total system energy is not guaranteed, making these systems inherently unsuitable for Hamiltonian descriptions.
Given this background, the formulation of collective motion in active matter typically involves direct descriptions of forces rather than using free energy.

\section{Related Work}\label{sec:relatedwork}

\paragraph{\TODO{System Identification methods:}} This section discusses methods for estimating governing laws from data, commonly referred to as system identification. System identification aims to estimate the equations of motion of a system from data. One well-known method for estimating differential equations is Sparse Identification of Nonlinear Dynamics (SINDy)~\cite{champion2019}. SINDy estimates nonlinear differential equations from data using LASSO regression to sparsely estimate the terms of the equations.
However, when applying this method to many-body systems with pairwise interactions, the number of parameters to estimate increases exponentially with the number of individuals, which raises computational costs and leads to potential instabilities. Other system identification methods use Bayesian optimization~\cite{bruckner2019, bruckner2020, bruckner2021}, e.g. representing the functions in the motion equations through basis function expansion, and estimating their coefficients via Bayesian optimization. This method has been successfully used in Vicsek model, reproducing its formation of orientational order.
However, the effectiveness of this approach for more complex systems, such as mixed-species systems, remains unclear. 
Deep learning approaches to analyzing collective motion have also been proposed~\cite{heras2019, ruiz-garcia2022, koyama2023}. 
\SKS{These incorporate Attention mechanisms to analyze behaviors limited to turning right or left~\cite{heras2019}, systems that limit the scope to Vicsek-type models to estimate orientational order parameters~\cite{ruiz-garcia2022}, or to estimate two-body interactions by expressing forces as neural networks~\cite{koyama2023}. Another possibility is to express Hamiltonians~\cite{Greydanus2019} or Lagrangians~\cite{cranmer2019lagrangian} as neural networks, in order to preserve known symmetries of the studied system. Some proposed including knowledge or assumptions about the physical context into the loss function~\cite{linka2022a}.
} 
\SKS{So far, none of these methods have been applied to systems in which multiple species interact with each other in unknown ways.}


\paragraph{\TODO{Integration of graph neural network with neural differential equations:}} 
\MU{
In this paper, we integrate GNN with neural ordinary differential equations (neural ODEs) to estimate interactions through evaluating equations of motion with dynamic edge structure. 
Most of the previous studies on integration of the two methods solve and train ODEs on a graph with fixed edges and consider changes in the edge weight~\cite{xhonneux2020, poli2021a, chen2023b, han2023, zhang2022b}.
This approach requires a fully or almost fully connected graph for collective motion where the adjacency changes drastically over time, and causes memory inefficiency.
On the other hand, some introduced a given time series of edge structure~\cite{pan2024}, making the system difficult to be extrapolated.
Thus we resolved these issues by setting a rule to define edges at each time point instead of explicitly giving graph structure time series.
}

\section{Method}\label{sec:method}

We aim to estimate the rules of motion for individual entities within collective motion data. This section presents the general framework that is used to generate the training data from a physical model for two numerical experiments, as well as the learning algorithm and neural network architecture.

\subsection{Physical model}
We represent the state of each entity \( i \) at time \( t \) as \( \mathcal{Z} \ni z^i(t) = (x^i(t), y^i) \in \mathcal{X} \times \mathcal{Y} \), where \( x^i \in \mathcal{X} \) denotes the variables described by the motion equations in the state space \( \mathcal{X} \subset \mathbb{R}^n \), and \( y^i \), a non-temporal variable, represents auxiliary attributes such as the type of each entity within the feature space \( \mathcal{Y} \).
We define a distance function \( d: \mathcal{Z} \times \mathcal{Z} \to \mathbb{R} \), and assume that entities \( i \) and \( j \) interact \TODO{at time \(t\) if \( d_{ij} (t) := d(z^i(t), z^j(t)) < d_0 \)}, where \( d_0 \in \mathbb{R} \) is a predefined threshold. 
The motion of each entity \( i\) is governed by the following Langevin equation:
\begin{equation}
dx^i = \left(F^{(1)}(z^i(t)) + \sum_{j \, \text{s.t.} \, d_{ij}(t)<d_0 \wedge j\neq i} F^{(2)}(z^i(t), z^j(t))\right) dt + \sigma dW^i(t),
\label{eq:eom}
\end{equation}
with given self-driven forces \( F^{(1)}: \mathcal{Z} \to \mathbb{R}^n \), forces due to interactions between pairs \( F^{(2)}: \mathcal{Z} \times \mathcal{Z} \to \mathbb{R}^n \), intensity of noise \( \sigma \in \mathbb{R} \), and a Wiener process \( W^i(t) \in \mathbb{R}^n \).
Given the initial state \( Z(t_0) \in \mathcal{Z} \) at time \( t_0 \in \mathbb{R} \), the state \( Z(t_0 + \Delta t) \in \mathcal{Z} \) after a time interval \( \Delta t \in \mathbb{R} \) can be determined by solving the motion equation using the operator \( S(F^{(1)}, F^{(2)}, \sigma, d, d_0): (Z(t_0), \Delta t) \mapsto Z(t_0 + \Delta t) \). Here,  $z^i(t_0+\Delta t) = (x^i(t_0+\Delta t), y^i)$ is defined by the following integral:
\begin{eqnarray}
x^i(t_0 + \Delta t) = x^i(t_0) + \int_{t_0}^{t_0 + \Delta t} dx^i(t),
\label{eq:integral}
\end{eqnarray}
which is computed numerically (\cref{sec:simulation_details}).
 \TODO{For \( \sigma = 0 \), we evaluated eq.\ref{eq:integral} with the neural ODE method~\cite{chen2019a, politorchdyn}, and for \( \sigma \neq 0 \) the neural stochastic differential equation (neural SDE) method~\cite{li2020scalable, kidger2021neuralsde}. }


\subsection{\TODO{Interpretation of the system as a graph}}
 
\TODO{The system at any time \( t \) can be represented as a directed graph \( G(t) = (V, E(t), Z(t)) \), where the set of entities represent the vertices \( V = \{1, 2, \ldots, N\} \), pairs of interacting entities constitute edges \( E(t)  = \{(i,j)\in V\times V | d_{ij}(t)<d_0 \wedge i\neq j\} \), and \( Z(t) = \{z^i(t) | i \in V\} \) is the set of states of all entities. Importantly, since the entities change states over time, the graph in general evolves over time. The forces \(F^{(1)}\) and \(F^{(2)}\) are evaluated on each vertex and each edge, respectively.}

\begin{figure}[ht]
 \centering
\includegraphics[width=\linewidth]{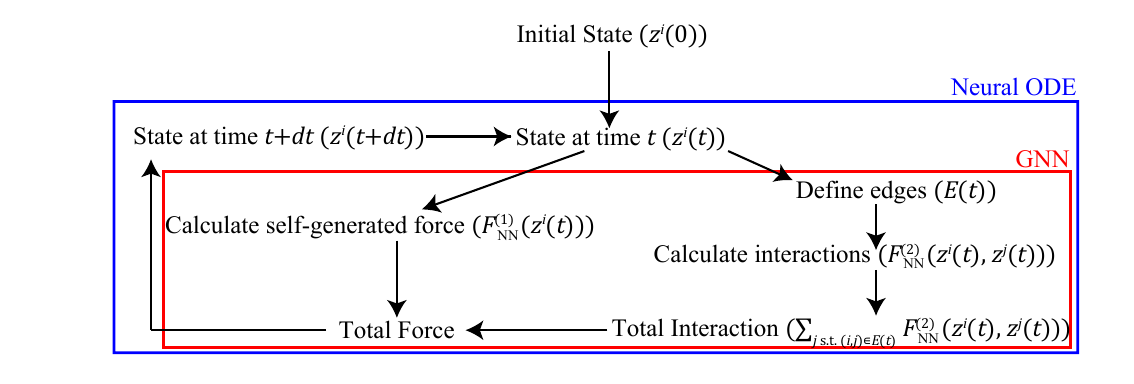}
\caption{\TODO{Schematic figure of the integration of GNN and neuralODE.}
\SKS{For details, see \cref{sec:simulation_details}.}}
 \label{fig:0}
\end{figure}

\subsection{\TODO{Solving the equations of motion of the graph through neural DEs}}
\label{sec:simulation_details}

\TODO{To obtain a time series of collective entities (eq. \ref{eq:integral}), we have to evaluate the equation of motion (eq. \ref{eq:eom}). 
We achieved this with a custom-made wrapper that connects neural ODE (SDE) to GNN (Figure \ref{fig:0}). 
The wrapper is equipped with the distance function \(d\) and the threshold \(d_0\) to define the edges \(E(t)\) at each time \(t\), the functions to calculate self-propulsion \(F^{(1)}\) and interaction \(F^{(2)}\), and the noise intensity \(\sigma\).
\SKS{The computation \(S(F^{(1)}, F^{(2)}, \sigma, d, d_0)\) is then run through the following steps:}}

\begin{enumerate}
\setcounter{enumi}{-1}
    \item \TODO{Create a graph object only with nodes \(V\) and set the static states \(\{y^i\}\) of the entities. Store this graph in the wrapper object. 
    Pass the initial state \(\{x^i(t_0)\}\) into the neural ODE (SDE) object (Figure \ref{fig:0} blue).}

    \item \TODO{The neural ODE (SDE) object passes the current state \(\{x^i(t)\}\) to the {\it forward} function of the wrapper object. }\label{step_ODEinitialize}

    \item \TODO{The {\it forward} function places the state values \(\{x^i(t)\}\) at the nodes of the graph. Subsequently, the graph is passed to GNN (Figure \ref{fig:0} red).}

    \item \TODO{GNN first defines and updates the edges \(E(t)\) using the distance function \(d\) and the threshold \(d_0\) stored in the wrapper. Next, it calculates the interactions \(F^{(2)}\) on the edges and adds them to the self-propulsion \(F^{(1)}\) on the nodes to return the forces to the ODE (SDE) object. If SDE, GNN also returns the noise intensity \(\sigma\) in the wrapper.}

    \item \TODO{The neural ODE (SDE) object updates the state \(\{x^i(t+dt)\}\). }\label{step_ODEupdate}

    \item \TODO{Repeat steps \ref{step_ODEinitialize}-\ref{step_ODEupdate} until the end of the time series.}
\end{enumerate}

\TODO{When inferring the forces \(F^{(1)}, F^{(2)}\) , we model them using neural networks (\cref{sec:learning}).}

\subsection{\TODO{Generation of training data}}

\TODO{To construct \(S_{\text{sim}} = S(F^{(1)}_{\text{sim}}, F^{(2)}_{\text{sim}}, \sigma_{\text{sim}}, d_{\text{sim}}, d_{0,\text{sim}})\),} distance functions \(d_{\text{sim}}\) were provided, suitable choices for the noise magnitude \(\sigma_{\text{sim}}\), threshold \(d_{0,\text{sim}}\) \TODO{and forms for the forces \(F^{(1)}_{\text{sim}}\), \(F^{(2)}_{\text{sim}}\) were specified (see section \ref{sec:Exp}). Thus defined \(S_{\text{sim}}\) was used to conduct simulations according to the motion equations, from initial states assigned uniformly at random \(x^i(0) \in \mathcal{X}_0\) , with }appropriate values for \(y^i \in \mathcal{Y}\).  The simulation was repeated \(M\) times, and the \(m\)-th result is denoted as \(Z^m_{\text{sim}}: t \mapsto S_{\text{sim}}(Z^m_{\text{sim}}(0), t)\) with the \(m\)-th initial condition \(Z^m_{\text{sim}}(0)\).
The combined data set \(Z_{\text{sim}} = \{Z^m_{\text{sim}} | m = 1, \ldots, M\}\) was then used to apply the subsequent learning algorithm, aiming to estimate \(F^{(1)}_{\text{sim}}\) and \(F^{(2)}_{\text{sim}}\).

\begin{figure}[h]
 \centering
\includegraphics[width=\linewidth]{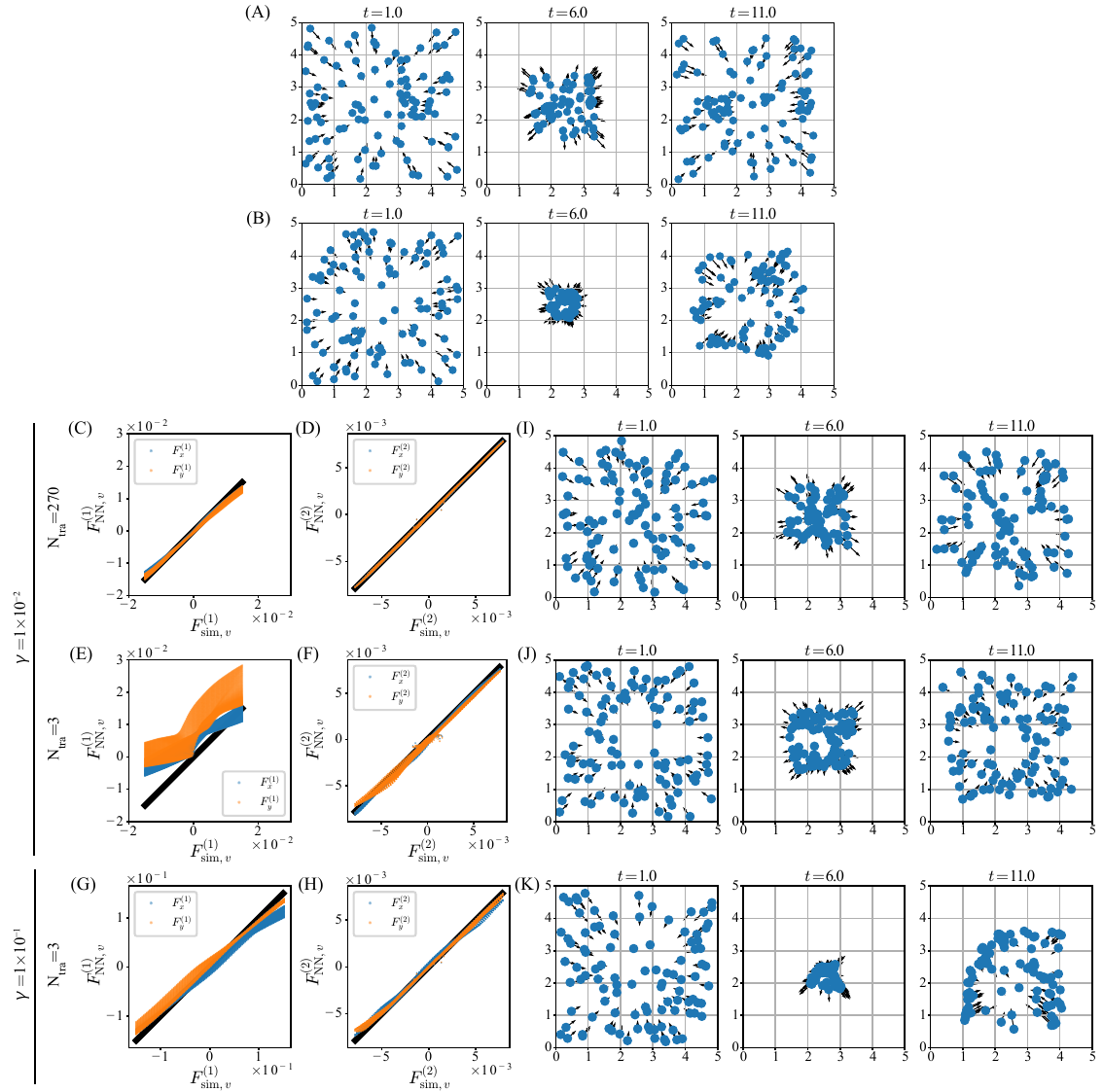}
\caption{Snapshots of the simulation results in the harmonic interaction model with friction constants. (A) The case with a friction constant of \(\gamma = 1 \times 10^{-2}\) and (B) with \(\gamma = 1 \times 10^{-1}\). Positions of individuals are indicated by blue circles, and velocities by black arrows.
(C-H) The functions estimated from the data with harmonic interaction plotted against the true values. The upper panel displays \(F^{(1)}\) and the lower panel \(F^{(2)}\) for different cases: (C-D) with \(\gamma = 1 \times 10^{-2}\) and \(N_\text{tra}=270\), (E-F) with \(\gamma = 1 \times 10^{-2}\) and \(N_\text{tra}=3\), and (G-H) with \(\gamma = 1 \times 10^{-1}\) and \(N_\text{tra}=3\). Blue and orange indicates x- or y- element $F$. A black line serves as a guide indicating where the estimated values equal the true values.
(I-K) Snapshots of the simulation results in the model estimated from the data with harmonic interaction. The panels represent different scenarios: (I) with \(\gamma = 1 \times 10^{-2}\) and \(N_\text{tra}=270\), (J) with \(\gamma = 1 \times 10^{-2}\) and \(N_\text{tra}=3\), and (K) with \(\gamma = 1 \times 10^{-1}\) and \(N_\text{tra}=3\).}
 \label{fig:1}
\end{figure}

\subsection{\TODO{Learning algorithm and neural network architecture for the forces}}
\label{sec:learning}
Subsequently, we developed a learning algorithm to estimate \(F^{(1)}_{\text{sim}}\) and \(F^{(2)}_{\text{sim}}\) using the results from the simulations conducted. 

The self-propulsion and interaction functions, \(F^{(1)}_{\text{NN}}\) and \(F^{(2)}_{\text{NN}}\), were modeled using neural networks. Unless otherwise specified, these functions consisted of a three-layer fully connected network followed by a scaling layer.
Each fully connected layer consisted of 128 nodes, with the Exponential Linear Unit (ELU) serving as the activation function. \MU{The scaling layer involved a scalar \(A\in\mathbb{R}\) and a vector \(B\in\mathbb{R}^n\) as learnable parameters}, and it transformed an input vector \(\alpha \in \mathbb{R}^n\) to \(e^A\alpha + B\). This configuration allowed the output of the fully connected network to be scaled appropriately, reflecting physical scales.

The parameters of fully connected networks and the scaling layer, collectively denoted \(\theta\), were optimized to approximate \(F^{(1)}_{\text{NN}}(\theta)\) and \(F^{(2)}_{\text{NN}}(\theta)\) to \(F^{(1)}_{\text{sim}}\) and \(F^{(2)}_{\text{sim}}\), respectively. To evaluate the deviation of \(F^{(1)}_{\text{NN}}(\theta)\) and \(F^{(2)}_{\text{NN}}(\theta)\), we first solved the motion equations using these functions.
\TODO{In practice, we constructed a wrapper object} \(S_{\text{NN}}(\theta) = S(F^{(1)}_{\text{NN}}(\theta), F^{(2)}_{\text{NN}}(\theta), 0, d_{\text{sim}}, d_{0,\text{NN}})\) and used it to perform simulations. Here, \(d_{0,\text{NN}}\) was a suitably set threshold, and for simplicity, \(\sigma_{\text{NN}} = 0\). In the simulation, the initial state was set as \TODO{a snapshot in training data} \(Z^m_{\text{sim}}(t_0)\), and we used this to compute the state \(\tau\) time units later as \(Z^m_{\text{NN}}(\theta; t_0+\tau) = S_{\text{NN}}(\theta)(Z^m_{\text{sim}}(t_0), \tau)\)\TODO{, which would be compared with a truth \(Z^m_{\text{sim}}(t_0 + \tau)\).}

\TODO{For each numerical experiment, a} loss function \(\mathcal{L}:\mathcal{Z}\times\mathcal{Z}\to\mathbb{R}\) was used to evaluate the discrepancy between the solution by \(\theta\) and the simulation data through \(\mathcal{L}^m(\theta;t_0) = \mathcal{L}(Z^m_{\text{sim}}(t_0+\tau), Z^m_{\text{NN}}(\theta; t_0+\tau))\). To optimize \(\theta\), the average \(\mathcal{L}(\theta)\) of this loss function over various \(m\) and \(t_0\) was minimized:
\begin{equation}
    \theta^* = \arg\min_{\theta} \mathcal{L}(\theta)
\end{equation}
The \(M\) simulation data sets were split into two parts, with \(M_{\text{tra}}\) sets used as training data and \(M_{\text{val}}\) sets used as validation data.
This optimization was conducted using the LAMB optimization algorithm~\cite{you2020}, leveraging gradient information with respect to \(\theta\).
The parameter \(\theta^*\) was evaluated using the minimum of the loss function calculated on the validation data set.

\section{Experiments} \label{sec:Exp}

\TODO{Here, we demonstrate the efficacy of the proposed method in} two numerical experiments.

\subsection{Underdamped Brownian Motion with Harmonic Interaction} \label{sec:harmonic}

\paragraph{\TODO{Training data:}}
First, to demonstrate that the self-propulsion and interaction forces could be estimated using our method, we performed simulations using a toy model. We used \(N=100\) particles, each with position \(r^i(t) \in \mathbb{R}^2\) and velocity \(v^i(t) \in \mathbb{R}^2\), such that \(z^i(t) = x^i(t) = (r^i(t), v^i(t))\) and \(y^i\) is empty.
The particles were connected by harmonic oscillators, and each velocity was subjected to a damping force, so the equations of motion read:
\begin{eqnarray}
    dr^i &=& v^i dt, \\
    dv^i &=& \left(-\gamma v^i - \sum_{j \text{s.t.} (i,j)\in E(t)} \nabla_{r^i} U(r^i-r^j)\right) dt + \sigma dW^i(t), \\
    U(r) &=& \frac{1}{2}k(|r|-r_c)^2,
\end{eqnarray}
where \(k \in \mathbb{R}\) is the strength of the interaction, \(r_c \in \mathbb{R}\) the natural spring length, and \(\gamma \in \mathbb{R}\) the friction coefficient. The distance function was \(d_{\text{sim}}(z^i, z^j) = |r^i - r^j|\), with a threshold set at \(d_{0,\text{sim}} = 5\).

In order to fit these terms into the aforementioned framework, the equations were reformulated as:
\begin{eqnarray}
    F^{(1)}_{\text{sim}}(z^i) &=& (F^{(1)}_{\text{sim}, r}(z^i), F^{(1)}_{\text{sim}, v}(z^i)) = (v^i, -\gamma v^i), \\
    F^{(2)}_{\text{sim}}(z^i, z^j) &=& (F^{(2)}_{\text{sim}, r}(z^i, z^j), F^{(2)}_{\text{sim}, v}(z^i, z^j)) = (0, -\nabla_{r^i} U(r^i-r^j)).
\end{eqnarray}
Here, \(k = 1 \times 10^{-3}\), \(r_c = 1\), \(\sigma = 1 \times 10^{-3}\), and \(\gamma\) was set to \(1 \times 10^{-2}\) or \(1 \times 10^{-1}\). The initial states \(x^i(0)\) were uniformly sampled from \(\mathcal{X}_0 = [0,5]^2 \times [0,1 \times 10^{-3}]^2\), and simulations were performed to obtain \(Z_{\text{sim}}\).

For these simulations, the Euler-Maruyama method with a time step of 0.1 was used, and data were collected at \(t = 0,1,\ldots,50\). The data showed that all particles cyclically gathered towards the center before dispersing (Figure \ref{fig:1}(A-B)). When \(\gamma = 1 \times 10^{-2}\), the amplitude of aggregation-dispersion was nearly constant (Figure \ref{fig:1}(A), Supplemental Movie S1), whereas for \(\gamma = 1 \times 10^{-1}\), the amplitude of aggregation-dispersion decreased (Figure \ref{fig:1}(B), Supplemental Movie S2).



\paragraph{Learning:}
With these data, we applied our proposed method to estimate \(F^{(1)}_{\text{sim}}\) and \(F^{(2)}_{\text{sim}}\). \(F^{(1)}_{\text{NN}}\) and \(F^{(2)}_{\text{NN}}\) were modeled as follows:
\begin{eqnarray}
    F^{(1)}_{\text{NN}}(z^i; \theta) &=& (v^i, F^{(1)}_{\text{NN}, v}(v^i; \theta)), \\
    F^{(2)}_{\text{NN}}(z^i, z^j; \theta) &=& (0, F^{(2)}_{\text{NN}, v}(r^j - r^i; \theta)).
\end{eqnarray}
\(F^{(1)}_{\text{NN}, v}\) and \(F^{(2)}_{\text{NN}, v}\) were each independent neural networks as described in \cref{sec:learning}. We evaluated the predictive accuracy for position and velocity as follows, using \(Z^m_{\text{NN}}(\theta; t_0+\tau)\) predicted for \(\tau = 10\) and the simulation data \(Z^m_{\text{sim}}(t_0)\):
\begin{eqnarray}
    \mathcal{L}_r(Z^m_{\text{sim}}(t_0+\tau), Z^m_{\text{NN}}(\theta; t_0+\tau)) &=& \frac{1}{N}\sum_{i\in V} |r^i_{\text{NN}}(\theta; t_0+\tau) -  r^i_{\text{sim}}(t_0+\tau)|^2, \\
    \mathcal{L}_v(Z^m_{\text{sim}}(t_0+\tau), Z^m_{\text{NN}}(\theta; t_0+\tau)) &=& \frac{1}{N}\sum_{i\in V} |v^i_{\text{NN}}(\theta; t_0+\tau) -  v^i_{\text{sim}}(t_0+\tau)|^2.
\end{eqnarray}
We sampled these metrics for 60 randomly selected pairs \((m, t_0)\). The total loss function \(\mathcal{L}(\theta)\) for one batch was defined as the dimensionless sum of these metrics, normalized by the variance in the simulation data:
\begin{eqnarray}
    \mathcal{L}_r(\theta) &=& \sum_{(m,t_0)} \mathcal{L}_r(Z^m_{\text{sim}}(t_0+\tau), Z^m_{\text{NN}}(\theta; t_0+\tau))/\text{Var}_{i,(m,t_0)}[r^i_{\text{sim}}(t_0+\tau)], \\
    \mathcal{L}_v(\theta) &=& \sum_{(m,t_0)} \mathcal{L}_v(Z^m_{\text{sim}}(t_0+\tau), Z^m_{\text{NN}}(\theta; t_0+\tau))/\text{Var}_{i,(m,t_0)}[v^i_{\text{sim}}(t_0+\tau)], \\
    \mathcal{L}(\theta) &=& \mathcal{L}_r(\theta) + \mathcal{L}_v(\theta).
\end{eqnarray}
To minimize the loss function, we manually searched for optimal hyperparameters for the LAMB optimizer. We determined that \(\beta_1 = 0.5\), \(\beta_2 = 0.4\), \(\epsilon = 1 \times 10^{-6}\), and a learning rate of \(1 \times 10^{-3}\) effectively minimized the loss function, although convergence was notably slow. It is important to note that we were unable to identify hyperparameters that would allow for both faster convergence and adequate estimation of the functional forms of \(F^{(1)}\) and \(F^{(2)}\).
As a result of minimizing this loss function, we optimized \(\theta\) to estimate \(F^{(1)}_{\text{NN}}\) and \(F^{(2)}_{\text{NN}}\) (Figure \ref{fig:1} (C-H)). 
When \(\gamma = 1 \times 10^{-2}\) and the training set size was \(M_{\text{tra}} = 270\) and validation set size \(M_{\text{val}} = 30\), after 300 epochs (15 days), \(F^{(1)}_{\text{NN}, v}\) and \(F^{(2)}_{\text{NN}, v}\) were observed to approximate \(F^{(1)}_{\text{sim}, v}\) and \(F^{(2)}_{\text{sim}, v}\) respectively (Figure \ref{fig:1}(C-D)). In contrast, with \(M_{\text{tra}} = 3\) and \(M_{\text{val}} = 3\), especially \(F^{(1)}_{\text{NN}, v}\) did not approximate \(F^{(1)}_{\text{sim}, v}\) even after 6000 epochs (4 days; Figure \ref{fig:1}(E-F)). However, under the same dataset size but with \(\gamma = 1 \times 10^{-1}\), \(F^{(1)}_{\text{NN}, v}\) and \(F^{(2)}_{\text{NN}, v}\) were confirmed to approximate \(F^{(1)}_{\text{sim}, v}\) and \(F^{(2)}_{\text{sim}, v}\) (Figure \ref{fig:1}(G-H)).

The accuracy of the estimation results is quantified in Supplemental Table S\ref{tab:S1}. For each particle or pair in the dataset \(Z_{\text{sim}}\), we calculated the forces \(F^{(1)}_v\) and \(F^{(2)}_v\) and computed the Mean Squared Error (MSE) and Mean Absolute Error (MAE). These errors were then normalized by the L1 or L2 norm of \(F^{(1)}_{\text{sim},v}\) and \(F^{(2)}_{\text{sim},v}\) respectively, to provide dimensionless measures of accuracy.
Due to the extensive computation time required for estimation, comprehensive statistics could not be gathered. Instead, we present the results of all trials to illustrate the trends in estimation accuracy. As observed, when \(M_{\text{tra}} = 3\) with a low friction constant \(\gamma = 1 \times 10^{-2}\), significant estimation errors occurred. Conversely, the estimation errors for the interaction force \(F^{(2)}_v\) were largely unaffected by \(\gamma\), suggesting that the estimation of interaction and friction are somewhat independent.
However, when \(M_{\text{tra}} = 270\), a slight improvement in the accuracy of interaction estimates was noted compared to the \(M_{\text{tra}} = 3\) scenario. This observation indicates that the accuracy of estimation depends on the number of data points, for the interaction as well as the friction.

Furthermore, to verify how well these estimates fit the training data, we conducted simulations from random initial values \(Z(0)\) sampled in a similar manner to training data creation,  and visualized the results \(S_{\text{NN}}(\theta^*)(Z(0),t)\) (Figure \ref{fig:1}(I-K); Supplemental Movies S3-5).
In all cases, the estimates were confirmed to adequately replicate the training data for all cases, even with \(\gamma = 1 \times 10^{-2}\) and \(M_{\text{tra}} = 3\). The ability to replicate training data suggests that friction had almost no effect in this case. However, this implies that the proposed method may not adequately estimate very weak effects.

\subsection{Mixed Species Collective Motion with Overdamped Self-propulsion}

\begin{figure}[h]
 \centering
\includegraphics[width=\linewidth]{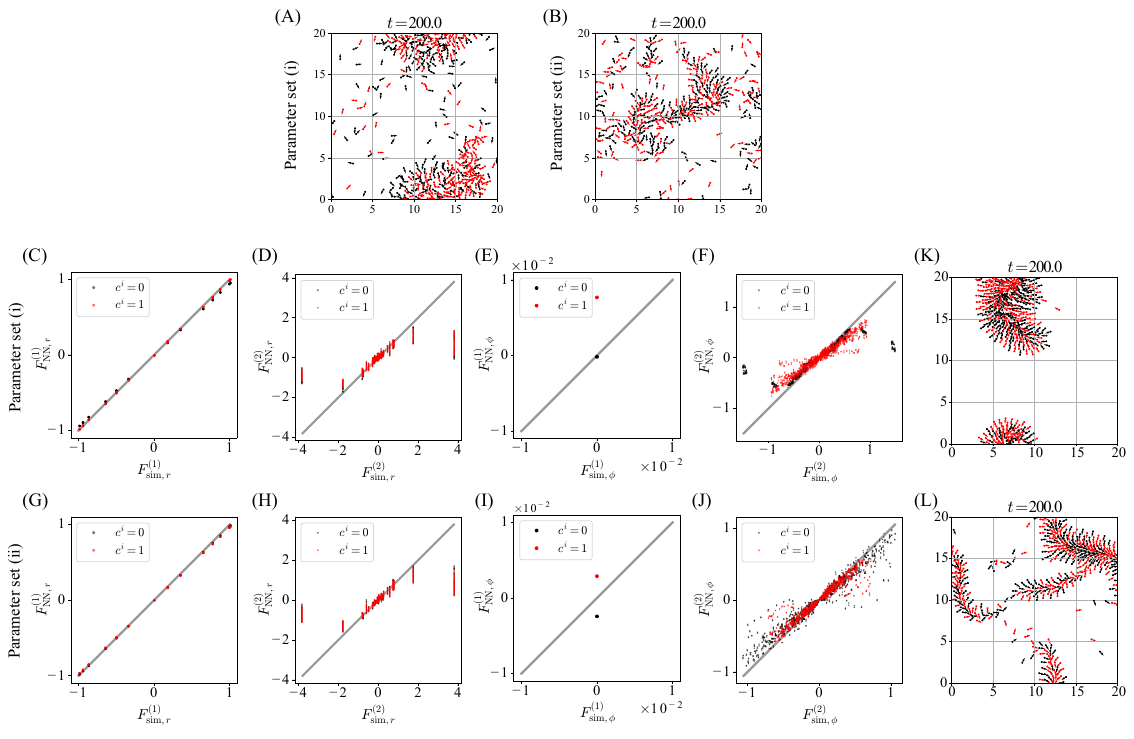}
\caption{Snapshots of the simulation results in the mixed-species model. Panel (A) displays \SKS{a representative snapshot of the training data for} the case with parameter set (i), and panel (B) with parameter set (ii) (see Supplemental Table S\ref{tab:S2}). Each individual is represented by an arrow located at their position and directed toward their polarity. Black arrows indicate individuals with \(c^i=0\), while red arrows represent those with \(c^i=1\).
(C-J) The functions estimated from the data with the mixed-species model plotted against the true values. The rows display different cases: (C-F) with parameter set (i), and (G-J) with parameter set (ii). A gray line serves as a guide indicating where the estimated values equal the true values.
(K-L) Snapshots of the simulation results in the model estimated from data with the mixed-species model. (K) The case with parameter set (i), and (L) with parameter set (ii). }
 \label{fig:2}
\end{figure}

Next, to test the proposed method in complex systems involving interactions among multiple species, we conducted simulations using a more complex model that emulates real collective movements. 

\paragraph{Training data:}
In this model, each of the \(N=400\) individuals has position \(r^i(t) \in [0, L]^2\), orientation \(\phi^i(t) \in [0, 2\pi]\), and species type \(c^i \in \{0,1\}\). Both \(r^i\) and \(\phi^i\) are subject to periodic boundary conditions, ensuring continuous and consistent movement dynamics across the defined space.
This defines \(x^i(t) = (r^i(t), \phi^i(t))\) and \(y^i = (c^i)\), constituting \(z^i(t) = (x^i(t), y^i)\).
The motion equations for individual \(i\) are described as follows:
\begin{eqnarray}
    dr^i &=& (v_0 p^i + \sum_{j\text{s.t.}(i,j)\in E(t)} \beta J_\text{eV}^{ij})  dt, \\
    d\phi^i &=& - \sum_{j\text{s.t.}(i,j)\in E(t)} \left( \alpha_\text{CF}(c^i) J_\text{CF}^{ij} + \alpha_\text{Ch}(c^i) J_\text{Ch}^{ij} \right) \left(r^{ij}\cdot p^i_\perp \right) dt + \sigma dW^i(t), \\
    J_\text{eV}^{ij} &=& \left(r_c^{-1} - |r^{ij}|^{-1}\right) r^{ij}, \\
    J_\text{CF}^{ij} &=& \frac{1}{2} (1-\frac{r^{ij}\cdot p^j}{|r^{ij}|}), \\
    J_\text{Ch}^{ij} &=& -\frac{r^{ij}\cdot p^i}{|r^{ij}|} K_1(\kappa |r^{ij}|), \\
    p^i &=& (\cos\phi^i, \sin\phi^i),\ \ p^i_\perp = (-\sin\phi^i, \cos\phi^i), \\
    r^{ij} &=& r^j - r^i \in [-L/2, L/2]^2,
\end{eqnarray}
with self-propulsion speed \(v_0 \geq 0\), the strength of the excluded volume interaction \(\beta \geq 0\), the strengths of contact following and chemotaxis \(\alpha_\text{CF}, \alpha_\text{Ch} \geq 0\) , respectively, the diffusion length of the chemoattractant \(\kappa \geq 0\), the noise strength \(\sigma \geq 0\), and the modified Bessel function of the second kind \(K_1\). 
The model was obtained by introducing a chemotaxis term into a preexisting model~\cite{Hiraiwa2020}, to make it more appropriate for cellular slime molds. The chemotaxis term assumes rapid diffusion of chemotactic substances secreted by each individual~\cite{Liebchen}.
We varied the sensitivity of contact following and chemotaxis depending on the species type, thus modeling a system where different species interact \SKS{non-reciprocally}. The interaction terms were adapted into our framework as follows:
\begin{eqnarray}
    F^{(1)}_{\text{sim}}(z^i) &=& (F^{(1)}_{\text{sim}, r}(z^i), F^{(1)}_{\text{sim}, \phi}(z^i)) \\
    &=& (v_0 p^i, 0), \\
    F^{(2)}_{\text{sim}}(z^i, z^j) &=& (F^{(2)}_{\text{sim}, r}(z^i, z^j), F^{(2)}_{\text{sim}, \phi}(z^i, z^j)) \\
    &=& (\beta J_\text{eV}^{ij}, -(\alpha_\text{CF}(c^i) J_\text{CF}^{ij} - \alpha_\text{Ch}(c^i) J_\text{Ch}^{ij}) \left(r^{ij}\cdot p^i_\perp \right)).
\end{eqnarray}
The distance function was \(d_{\text{sim}}(z^i, z^j) = |r^{ij}|\), with periodic boundary conditions of width \(L = 20\) and a threshold \(d_{0,\text{sim}} = 1\). The parameters were \(v_0 = 1\), \(\beta = 1\), \(\kappa = 0.5\), and \(\sigma = \sqrt{0.2}\), with 200 individuals of each species \(c^i = 0, 1\). Parameter dependencies based on \(c^i\) are shown in Supplemental Table S\ref{tab:S2}.

The initial states \(x^i(0)\) were uniformly sampled from \([0, L]^2 \times [0, 2\pi]\), and simulations were carried out to generate \(Z_{\text{sim}}\) using the Heun method with a timestep of 0.1, collecting data for \(t = 0, 1, \ldots, 300\). Simulation results for parameter set (i) showed the formation of mixed-species circular clusters exhibiting rotational behavior (Figure \ref{fig:2}(A), Supplemental Movie S6). Conversely, results for parameter set (ii) formed centipede-like clusters of mixed species that moved translationally in local alignments (Figure \ref{fig:2}(B), Supplemental Movie S7).



\paragraph{Learning:}
In order to estimate \(F^{(1)}_{\text{sim}}\) and \(F^{(2)}_{\text{sim}}\), we modeled \(F^{(1)}_{\text{NN}}\) and \(F^{(2)}_{\text{NN}}\) as follows:
\begin{eqnarray}
    F^{(1)}_{\text{NN}}(z^i; \theta) &=& (\mathcal{R}(\phi^i)F^{(1)}_{\text{NN}, r}(c^i; \theta), F^{(1)}_{\text{NN}, \phi}(c^i; \theta)), \\
    F^{(2)}_{\text{NN}}(z^i, z^j; \theta) &=& \left( \begin{array}{c} \mathcal{R}(\phi^i)F^{(2)}_{\text{NN}, r}(\mathcal{R}(-\phi^i)r^{ij}, \phi^j - \phi^i, c^i, c^j; \theta) \\
    F^{(2)}_{\text{NN}, \phi}(\mathcal{R}(-\phi^i)r^{ij}, \phi^j - \phi^i, c^i, c^j; \theta) \end{array} \right)^T.
\end{eqnarray}
Here, \(F^{(1)}_{\text{NN}, r}, F^{(1)}_{\text{NN}, \phi}, F^{(2)}_{\text{NN}, r}, F^{(2)}_{\text{NN}, \phi}\) are independent neural networks with their respective parameters, and \(\mathcal{R}(\phi^i)\) is the rotation matrix for \(\phi^i\), which converts all angles into relative angles with respect to \(\phi^i\), thereby maintaining the model's invariance to coordinate rotation. Additionally, when inputting the categorical variable \(c^i\) into the neural network, it is transformed into a two-dimensional vector via an embedding layer.
The predictions \(Z^m_{\text{NN}}(\theta; t_0+\tau)\) made by this neural network for \(\tau = 1\) and the simulation data \(Z^m_{\text{sim}}(t_0)\) defined the prediction errors for position and orientation as follows:
\begin{eqnarray}
    \mathcal{L}_r(Z^m_{\text{sim}}(t_0+\tau), Z^m_{\text{NN}}(\theta; t_0+\tau)) &=& \frac{1}{N}\sum_{i\in V} d(r^i_{\text{NN}}(\theta; t_0+\tau), r^i_{\text{sim}}(t_0+\tau))^2, \\
    \mathcal{L}_\phi(Z^m_{\text{sim}}(t_0+\tau), Z^m_{\text{NN}}(\theta; t_0+\tau)) &=& \frac{1}{N}\sum_{i\in V} \left(1 - \cos(\phi^i_{\text{NN}}(\theta; t_0+\tau) - \phi^i_{\text{sim}}(t_0+\tau))\right).
\end{eqnarray}
Predictions were computed using the Euler-Maruyama method with a timestep of 0.1. These metrics were sampled for 60 randomly selected pairs \((m, t_0)\) and each was normalized by the variance in the simulation data to compose the loss function for one batch:
\begin{eqnarray}
    \mathcal{L}_r(\theta) &=& \sum_{(m,t_0)} \mathcal{L}_r(Z^m_{\text{sim}}(t_0+\tau), Z^m_{\text{NN}}(\theta; t_0))/\text{Var}_{i,(m,t_0)}[r^i_{\text{sim}}(t_0+\tau)], \\
    \mathcal{L}_\phi(\theta) &=& \sum_{(m,t_0)} \mathcal{L}_\phi(Z^m_{\text{sim}}(t_0+\tau), Z^m_{\text{NN}}(\theta; t_0))/\sum_{i,(m,t_0)}\left(1-\cos(\phi^i_{\text{sim}}(t_0+\tau))\right), \\
    \mathcal{L}(\theta) &=& \mathcal{L}_r(\theta) + \mathcal{L}_\phi(\theta).
\end{eqnarray}
To minimizing this loss function, we tried the same hyperparameters as those used for the harmonic interaction model, and found we optimized \(\theta\) to estimate \(F^{(1)}_{\text{NN}}\) and \(F^{(2)}_{\text{NN}}\) (Figure \ref{fig:2}(C-J)). 
All experiments were conducted with \(M_{\text{tra}} = 3, M_{\text{val}} = 3\). After 20,000 epochs (25 days), \(F^{(1)}_{\text{NN}, r}, F^{(1)}_{\text{NN}, \phi}, F^{(2)}_{\text{NN}, r}, F^{(2)}_{\text{NN}, \phi}\) were found to approximate \(F^{(1)}_{\text{sim}, r}, F^{(1)}_{\text{sim}, \phi}, F^{(2)}_{\text{sim}, r}, F^{(2)}_{\text{sim}, \phi}\) closely (Figure \ref{fig:2}(C-J)).
To further verify the fit of our estimation results with the training data, we sampled random initial values \(Z(0)\) similarly to the training data creation process and conducted simulations to visualize the outcomes \(S_{\text{NN}}(\theta^*)(Z(0),t)\) (Figure \ref{fig:2}(K-L); Supplemental Movies S8-9). The results confirmed that all estimates adequately reproduced the training data.

To further test the predictive power of the model trained on the mixed state, we performed simulations where all individuals were of the same type. 
For initial conditions \(Z_c(0)\) where all \(N=400\) individuals shared the same type \(c^i=c\), either \(c=0\) or \(c=1\), we performed simulations \MU{using the aforementioned model \(S_{\text{sim}}\) and the neural network \(S_{\text{NN}}(\theta^*)\) trained above (Supplemental Figure S\ref{fig:S1}).}
For parameter set (i) with \(c=0\), both \(S_{\text{sim}}(Z_0(0),t)\) and \(S_{\text{NN}}(\theta^*)(Z_0(0),t)\) showed the formation of two clusters that exhibited rotational movements (Supplemental Figure S\ref{fig:S1}(A-B)). 
Conversely, for parameter set (i) with \(c=1\) and parameter set (ii) with \(c=0,1\), both \(S_{\text{sim}}(Z_0(0),t)\) and \(S_{\text{NN}}(\theta^*)(Z_0(0),t)\) formed centipede-like clusters that moved translationally in alignment with their local direction (Supplemental Figure S\ref{fig:S1}(C-H)).
\MU{These results suggest that the interaction of each type was approximated accurately enough to explain the dynamics of training data.}

As a negative control, we conducted simulations where both types of individuals shared the same parameters, as in parameter sets (iii)-(v), and \MU{trained the neural networks} for each case (Supplemental Figure S\ref{fig:S2}). 
In all instances, it was confirmed that \(F^{(1)}_{\text{NN}, r}, F^{(1)}_{\text{NN}, \phi}, F^{(2)}_{\text{NN}, r}, F^{(2)}_{\text{NN}, \phi}\) closely approximated \(F^{(1)}_{\text{sim}, r}, F^{(1)}_{\text{sim}, \phi}, F^{(2)}_{\text{sim}, r}, F^{(2)}_{\text{sim}, \phi}\). 
Notably, these estimation results showed minimal dependency on \(c^i\) in \(F^{(1)}_{\text{NN}}(z^i; \theta^*), F^{(2)}_{\text{NN}}(z^i, z^j; \theta^*)\). 
This outcome suggests that the species dependency of interactions estimated by the proposed method in mixed assemblies is based on actual data, rather than being an artifact of the methodology.

Finally, to summarize the results above, we quantified the estimation accuracy for each trial, similar to the methods described in \cref{sec:harmonic} (Supplemental Table S\ref{tab:S3}). 
To note, since \(F^{(1)}_{\text{sim},\phi} = 0\), the MSE and MAE for \(F^{(1)}_{\phi}\) are not normalized. 
Our results demonstrate that \(F^{(1)}_r\) is estimated with high accuracy across all conditions.
Also, the estimation of \(F^{(1)}_{\phi}\) is  maintained at sufficiently small magnitudes relative to the order of magnitude (\(10^0\)) typically seen with \(F^{(2)}_\phi\) (Figure \ref{fig:2}(F,J)).
Regarding the interactions, both \(F^{(2)}_r\) and \(F^{(2)}_\phi\) exhibited somewhat higher errors, approximately of the order of \(10^{-1}\).
These errors likely stem from the high dimensionality of the inputs to the \(F^{(2)}_{\text{NN}}\), which may prevent the neural network from achieving sufficient learning, or from the training process neglecting the noise.

\section{Conclusion}\label{sec:conclusion}

In this study, we proposed a novel method for estimating interactions among individuals within models of collective motion. This method combines dynamic GNNs with neural ODEs to estimate interactions among individuals. We demonstrated that this method \SKS{was able to} estimate interactions in both simple models and complex mixed-species collective motion models.
In models with simple harmonic interactions, the proposed method \SKS{identified the relevant} interactions effectively. 
\SKS{Importantly, the method successfully inferred the non-reciprocal interactions between the different species} in the complex mixed-species \SKS{model for the collective dynamics of slime molds}.
\MU{For the latter case, \SKS{the continuous updating of the} edge structures in our approach substantially reduces memory \SKS{needs. In our simulations of 400 bodies} with \SKS{an} edge density of 2\%, the required memory of 200 GB for \SKS{a} fully connected graph is reduced to 30GB \SKS{thus making them} feasible on \SKS{off-the-shelf GPUs}. }
Nevertheless, we should note that this approach is limited due to the long time required for estimation. Moreover, the method currently estimates deterministic motion equations and only considers pairwise interactions, thereby not accommodating interactions among three or more bodies or the effects of noise. Future research is expected to extend this approach to estimate more general interactions and develop methods for stochastic motion equations. Applying the present methods to real data in systems such as immune cells should help clarify the complex rules behind their migration stategies.

\section*{\TODO{Acknowledgements}}
\SKS{This research was supported by JSPS KAKENHI Grants JP22H05673 (awarded to MU), JP22H04841 (SKS), JP22K14012 (SKS), JP19H05799 (TJK), JP19H05801 (SS) and JP19KK0282 (SS), as well as JST CREST Grants JPMJCR2011 (TJK) and JPMJCR1923 (SS).
\SKS{
We acknowledge the use of ChatGPT and GitHub Copilot for their assistance in swiftly articulating ideas in clear English. Passages written with the help of these tools were reviewed and rewritten by the authors.
}
}

\bibliographystyle{plain}
\bibliography{main}

\newpage
\appendix

\setcounter{figure}{0}
\setcounter{table}{0}
\captionsetup[figure]{labelformat=suppfigure}
\captionsetup[table]{labelformat=supptable}

\section{Supplemental Material}

\subsection*{Computational Resources}
Detailed specifications of the computational setup are provided here to ensure reproducibility. We employed the Dell Precision 7920 Tower, which comprises of 64 GB RAM, 2 Intel Xeon CPUs, and 2 NVIDIA RTX A6000 GPUs. The system operates under Linux Ubuntu 22.04, with all neural network training processes managed through pyenv with python 3.8.13 to maintain environment consistency. To use the GPUs in the computation, we used CUDA 11.5 and PyTorch 1.13.1.

\subsection*{Existing Assets}
To construct our framework, we utilized the Deep Graph Library 1.1.1 (\url{https://www.dgl.ai/})\cite{wang2020c}, TorchSDE 0.2.5 (\url{https://github.com/google-research/torchsde})\cite{li2020scalable,kidger2021neuralsde}, TorchDyn 1.0.4 (\url{https://github.com/DiffEqML/torchdyn})\cite{politorchdyn}, and torch-optimizer 0.3.0 (\url{https://github.com/jettify/pytorch-optimizer})\cite{Novik_torchoptimizers}, all of which are publicly available under the MIT License and Apache License 2.0.

\subsection*{Supplemental Tables}

\begin{table}[ht]
\centering
\begin{tabular}{c|c|c|c|c|c|c}
\toprule
\textbf{Trial ID} & \textbf{\(\gamma\)} & \textbf{\(M_\text{tra}\)} & \textbf{\(\text{MSE}(F^{(1)}_v)\)} & \textbf{\(\text{MAE}(F^{(1)}_v)\)} & \textbf{\(\text{MSE}(F^{(2)}_v)\)} & \textbf{\(\text{MAE}(F^{(2)}_v)\)} \\
\midrule
1 & \(1\times10^{-2}\) & 270 & \(3.6\times10^{-3}\) & \(5.9\times10^{-2}\) & \(5.1\times10^{-3}\) & \(4.8\times10^{-2}\) \\
  \hline
2 & \(1\times10^{-2}\) & 3 & \(1.7\) & \(1.4\) & \(2.2\times10^{-2}\) & \(1.4\times10^{-1}\) \\
3 & \(1\times10^{-2}\) & 3 & \(9.4\) & \(3.7\) & \(2.2\times10^{-2}\) & \(1.3\times10^{-1}\) \\
  \hline
4 & \(1\times10^{-1}\) & 3 & \(1.2\times10^{-2}\) & \(1.7\times10^{-2}\) & \(2.2\times10^{-2}\) & \(1.0\times10^{-1}\) \\
5 & \(1\times10^{-1}\) & 3 & \(5.4\times10^{-2}\) & \(3.8\times10^{-2}\) & \(2.3\times10^{-2}\) & \(1.1\times10^{-1}\) \\
\bottomrule
\end{tabular}
\caption{Normalized MSE and MAE of the estimation of \(F^{(1)}\) and \(F^{(2)}\) for the harmonic interaction model. Each row indicates an independent experiment.}
\label{tab:S1}
\end{table}

\begin{table}[ht]
\centering
\begin{tabular}{c|c|c|c|c}
\toprule
\textbf{Parameter set} & \textbf{\(\alpha_{CF}(0)\)} & \textbf{\(\alpha_{CF}(1)\)} & \textbf{\(\alpha_{Ch}(0)\)} & \textbf{\(\alpha_{Ch}(1)\)}  \\
\midrule
(i) & 0.1 & 0.9 & 2.0 & 0.2 \\
  \hline
(ii) & 0.9 & 0.5 & 0.5 & 0.5 \\
  \hline
(iii) & 0.9 & 0.9 & 0.2 & 0.2 \\
  \hline
(iv) & 0.9 & 0.9 & 0.5 & 0.5 \\
  \hline
(v) & 0.1 & 0.1 & 2.0 & 2.0 \\
\bottomrule
\end{tabular}
\caption{Parameter sets used in the simulations of the mixed-species model.}
\label{tab:S2}
\end{table}

\begin{table}[ht]
\begin{adjustbox}{max width=1\textwidth}
\centering
\begin{tabular}{c|c|c|c|c|c|c|c|c|c}
\toprule
\textbf{Trial ID} & \textbf{Parameter set} & \textbf{\(\text{MSE}(F^{(1)}_r)\)} & \textbf{\(\text{MAE}(F^{(1)}_r)\)} & \textbf{\(\text{MSE}(F^{(1)}_\phi)\)} & \textbf{\(\text{MAE}(F^{(1)}_\phi)\)} & \textbf{\(\text{MSE}(F^{(2)}_r)\)} & \textbf{\(\text{MAE}(F^{(2)}_r)\)} & \textbf{\(\text{MSE}(F^{(2)}_\phi)\)} & \textbf{\(\text{MAE}(F^{(2)}_\phi)\)} \\
\midrule
1 & (i) & \(1.5\times10^{-3}\) & \(3.7\times10^{-2}\) & \(4.5\times10^{-7}\) & \(6.7\times10^{-4}\) & \(5.6\times10^{-1}\) & \(2.7\times10^{-1}\) & \(2.4\times10^{-1}\) & \(2.8\times10^{-1}\) \\
2 & (i) & \(1.7\times10^{-3}\) & \(3.5\times10^{-2}\) & \(3.0\times10^{-7}\) & \(3.9\times10^{-3}\) & \(9.9\times10^{-1}\) & \(3.0\times10^{-1}\) & \(1.6\times10^{-1}\) & \(2.8\times10^{-1}\) \\
  \hline
3 & (ii) & \(2.0\times10^{-3}\) & \(4.2\times10^{-2}\) & \(3.8\times10^{-6}\) & \(1.7\times10^{-3}\) & \(1.9\times10^{-1}\) & \(1.6\times10^{-1}\) & \(5.4\times10^{-2}\) & \(2.0\times10^{-1}\) \\
4 & (ii) & \(5.8\times10^{-4}\) & \(2.3\times10^{-2}\) & \(7.1\times10^{-6}\) & \(2.7\times10^{-3}\) & \(2.7\times10^{-1}\) & \(1.7\times10^{-1}\) & \(5.8\times10^{-2}\) & \(2.1\times10^{-1}\) \\
  \hline
5 & (iii) & \(1.0\times10^{-3}\) & \(3.1\times10^{-2}\) & \(2.3\times10^{-6}\) & \(1.1\times10^{-3}\) & \(4.2\times10^{-1}\) & \(1.9\times10^{-1}\) & \(6.8\times10^{-2}\) & \(2.5\times10^{-1}\) \\
  \hline
6 & (iv) & \(1.6\times10^{-4}\) & \(1.3\times10^{-2}\) & \(3.2\times10^{-5}\) & \(5.0\times10^{-3}\) & \(3.7\times10^{-1}\) & \(1.7\times10^{-1}\) & \(7.3\times10^{-2}\) & \(2.4\times10^{-1}\) \\
  \hline
7 & (v) & \(2.3\times10^{-3}\) & \(4.8\times10^{-2}\) & \(3.9\times10^{-6}\) & \(1.9\times10^{-3}\) & \(2.8\times10^{-1}\) & \(1.9\times10^{-1}\) & \(8.8\times10^{-2}\) & \(1.6\times10^{-1}\) \\
8 & (v) & \(6.4\times10^{-3}\) & \(8.0\times10^{-2}\) & \(3.1\times10^{-5}\) & \(5.4\times10^{-3}\) & \(3.9\times10^{-1}\) & \(2.3\times10^{-1}\) & \(1.4\times10^{-1}\) & \(2.0\times10^{-1}\) \\
\bottomrule
\end{tabular}
\end{adjustbox}
\caption{Normalized MSE and MAE of the estimation of \(F^{(1)}\) and \(F^{(2)}\) for the mixed-species model. Each row indicates an independent experiment. To note, \textbf{\(\text{MSE}(F^{(1)}_\phi)\)} and \textbf{\(\text{MAE}(F^{(1)}_\phi)\)} are not normalized since the true value is zero in any case.}
\label{tab:S3}
\end{table}

\newpage

\subsection*{Supplemental figures}

 \begin{figure}[ht]
  \centering
 \includegraphics[width=\linewidth]{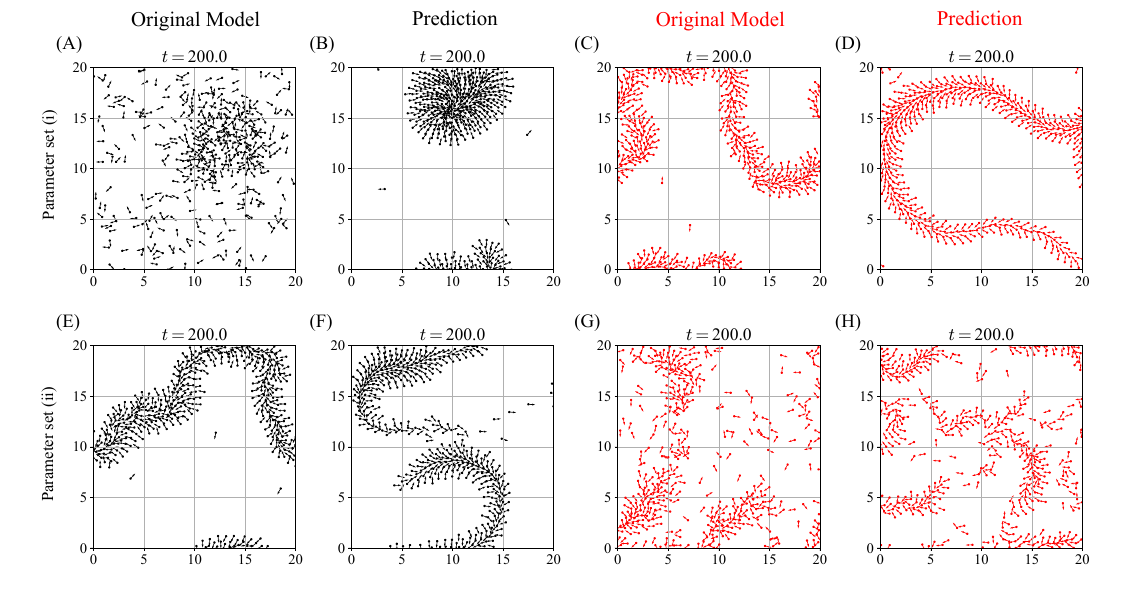}
 \caption{Snapshots of the simulation results in the mixed-species model and the estimated model. Panels (A, C, E, G) depict results from the mixed-species model, and panels (B, D, F, H) from the estimated model. The upper rows (A-D) represent simulations with parameter set (i), and the lower rows (E-H) with parameter set (ii). Panels (A, B, E, F) include individuals with \(c^i=0\) and panels (C, D, G, H) with \(c^i=1\). Colors correspond to those used in Figure \ref{fig:2}.}
  \label{fig:S1}
 \end{figure}

 \begin{figure}[ht]
  \centering
\includegraphics[width=\linewidth]{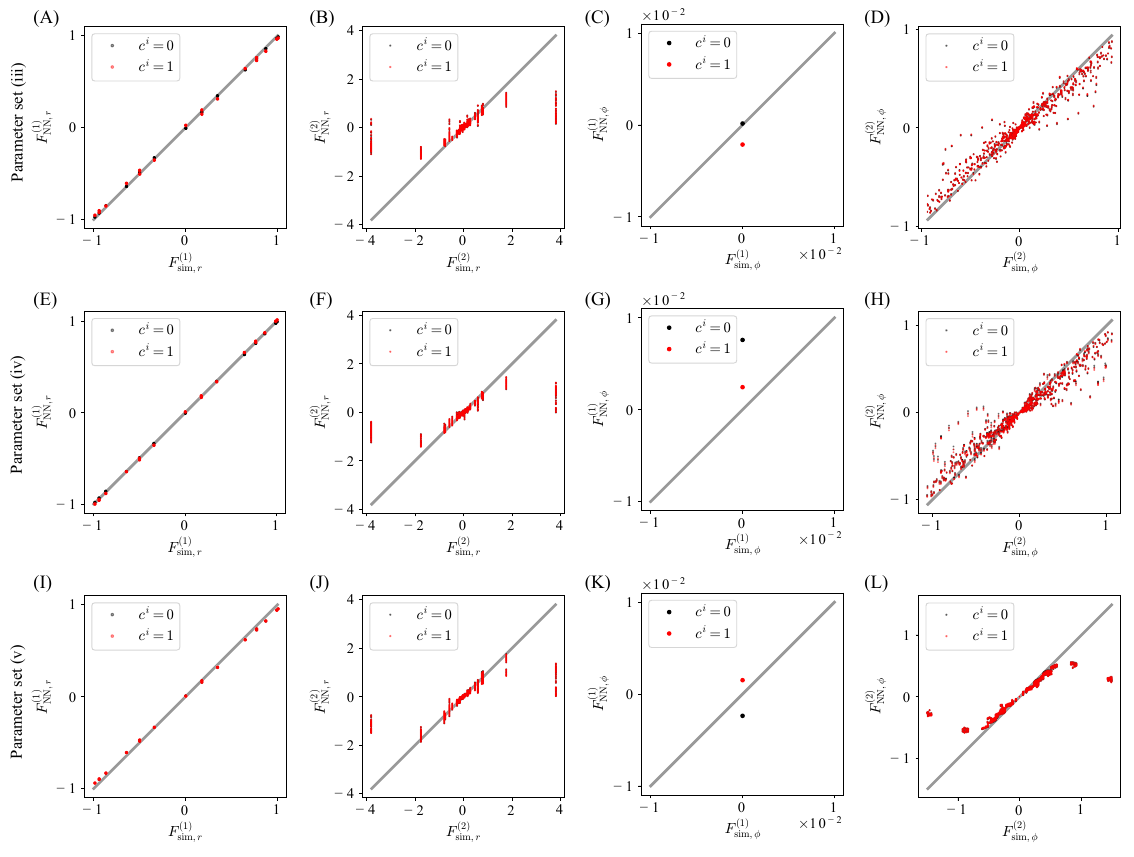}
 \caption{The functions estimated from data with the mixed-species model, without dependency on species type, plotted against the true values. The rows display different cases: (A-D) with parameter set (iii), (E-H) with parameter set (iv), and (I-L) with parameter set (v). A gray line serves as a guide to indicate where the estimated values equal the true values.}
  \label{fig:S2}
 \end{figure}
 
 \newpage

\subsection*{Supplemental Movies}

\textbf{Supplemental Movie S1:} 

A visualization of the simulation based on the harmonic interaction model with friction constant \(\gamma = 1 \times 10^{-2}\).

\textbf{Supplemental Movie S2:} 

A visualization of the simulation based on the harmonic interaction model with friction constant \(\gamma = 1 \times 10^{-1}\).

\textbf{Supplemental Movie S3:} 

The results of simulations using the estimated functions \(F^{(1)}_{\text{NN}}\) and \(F^{(2)}_{\text{NN}}\), which were trained using \(N_\text{tra}=270\) data from the harmonic interaction model with \(\gamma = 1 \times 10^{-2}\).

\textbf{Supplemental Movie S4:} 

The results of simulations using the estimated functions \(F^{(1)}_{\text{NN}}\) and \(F^{(2)}_{\text{NN}}\), which were trained using \(N_\text{tra}=3\) data from the harmonic interaction model with \(\gamma = 1 \times 10^{-2}\).

\textbf{Supplemental Movie S5:} 

The results of simulations using the estimated functions \(F^{(1)}_{\text{NN}}\) and \(F^{(2)}_{\text{NN}}\), which were trained using \(N_\text{tra}=3\) data from the harmonic interaction model with \(\gamma = 1 \times 10^{-1}\).

\textbf{Supplemental Movie S6:} 

A visualization of the simulation based on the mixed-species model with parameter set (i).

\textbf{Supplemental Movie S7:} 

A visualization of the simulation based on the mixed-species model with parameter set (ii).

\textbf{Supplemental Movie S8:} 

The results of simulations using the estimated functions \(F^{(1)}_{\text{NN}}\) and \(F^{(2)}_{\text{NN}}\), which were trained using \(N_\text{tra}=3\) data from the mixed-species model with parameter set (i).

\textbf{Supplemental Movie S9:} 

The results of simulations using the estimated functions \(F^{(1)}_{\text{NN}}\) and \(F^{(2)}_{\text{NN}}\), which were trained using \(N_\text{tra}=3\) data from the mixed-species model with parameter set (ii).


\end{document}